\documentclass[aps,twocolumn,showpacs,floats,pra]{revtex4}

\usepackage{graphicx}
\usepackage{dcolumn}
\usepackage{bm}
\usepackage{color}
\usepackage{subfigure}
\usepackage{amsmath}
\usepackage[normalem]{ulem}
\usepackage[colorlinks,bookmarks=false,citecolor=blue,linkcolor=red,urlcolor=blue]{hyperref}

\newcommand{\excitationOp}{\delta U}
\newcommand{\deltau}{\delta u}
\newcommand{\rv}{{\bf r}}
\newcommand{\qv}{{\bf q}}
\newcommand{\sumprime}{\sideset{}{'}\sum}
\newcommand{\diff}[1]{\mathrm{d} #1}
\newcommand{\hamiltonian}{{\mathcal{H}}}
\newcommand{\potential}{V}
\newcommand{\kv}{{\bf k}}

\begin{document}

\title{Gapped spectrum in pair-superfluid bosons}

\author{G. E. Astrakharchik$^1$, R. E. Zillich$^2$, F. Mazzanti$^1$, and J. Boronat$^1$}

\affiliation{$^1$ Departament de F\'{\i}sica, Universitat Polit\`ecnica de Catalunya, Campus Nord B4-B5, E-08034 Barcelona, Spain}

\affiliation{$^2$ Institut f\"ur Theoretische Physik, Johannes--Kepler Universit\"at, Altenbergerstr. 69, 4040 Linz, Austria}

\begin{abstract}
We study the ground state of a bilayer system of dipolar bosons with dipoles oriented by an external field perpendicularly to the two parallel planes.
By decreasing the interlayer distance, for a fixed value of the strength of the dipolar interaction, the system undergoes a quantum phase transition from an atomic to a pair superfluid.
We investigate the excitation spectrum across this transition by using microscopic approaches.
Quantum Monte Carlo methods are employed to obtain the static structure factors and intermediate scattering functions in imaginary time.
The dynamic response is calculated using both the correlated basis functions (CBF) method and the approximate inversion of the Laplace transform of the quantum Monte Carlo imaginary time data.
In the atomic phase, both density and spin excitations are gapless.
However, in the pair-superfluid phase a gap opens in the excitation energy of the spin mode.
For small separation between layers, the minimal spin excitation energy equals the binding energy of a dimer and is twice the gap value.
\end{abstract}

\pacs{05.30.Fk, 03.75.Hh, 03.75.Ss}

\maketitle

\section{Introduction}

The study of quantum dipolar gases has attracted much experimental and theoretical interest in the last decade, since pioneering works where trapped clouds of $^{52}$Cr atoms, brought close to a broad Feshbach resonance, revealed clear evidences of condensate deformation~\cite{Greismaier_2005,Lahaye_2007}.
After these initial experiments, new perspectives with other species with much larger dipolar moments were explored.
In this line, polar molecules of and Rb~\cite{Ospelkaus_2010}, or Cs and Rb~\cite{Kerman_2004}, with much stronger dipolar interactions, seemed to be the optimal candidates.
However, it turned out to be very difficult to keep them in the quantum degeneracy limit due to three-body losses and chemical reactions.
Still, an important progress has been recently achieved with NaK~\cite{Park_2015} and NaRb~\cite{Guo_2016} molecules.
In much the same way, new magnetic dipolar condensates of Dy~\cite{Lu_2011} and Er~\cite{Aikawa_2012} species have been produced, allowing for a much cleaner measurement of quantum dipolar physics due to the absence of most of the problems found when dealing with polar molecules.

From the theoretical side, the anisotropic and long-ranged character of the dipolar interaction makes these systems unique, exhibiting new features like $p$-wave superfluidity in two-dimensional Fermi gases~\cite{Brunn_2008} or roton instabilities~\cite{Santos_2003, Macia_2012}, that enrich the phase diagram when compared with other condensed matter systems governed by Van der Waals forces.
More recently, the formation of solid structures of droplets of trapped dipolar bosons brought to the regime of mean-field collapse has also attracted much attention, both from the experimental~\cite{Kadau_2016,Ferrier-Barbut_2016} and theoretical~\cite{Bisset_2015,Blakie_2016,Wachtler_2016,Kui-Tian_2016,Macia_2016} points of view.
An appealing setup which permits to avoid the collapse of the system is the bilayer, or even the multilayer.
For instance, in the case of Fermi dipoles it has been shown that the bilayer geometry produces a non-zero superfluid signal that, depending on the interlayer distance, is due either to BCS pairs or to tightly bound molecules in a BEC state~\cite{Pikovski_2010,Matveeva_2014}.
The bosonic counterpart of this problem has also been studied, revealing for the first time a homogeneous Bose system that undergoes a quantum phase transition from a single-particle to a pair superfluid with decreasing interlayer distance~\cite{Macia_2014}.

The interlayer separation in a bilayer (or multilayer) geometry introduces a potential barrier that helps to stabilize the system, reducing the effects induced by three-body loses, a fact that can be particularly relevant in the case of polar molecules~\cite{deMiranda_2011}.
This allows a tunable transition to a superfluid of dimers in the Bose case, strongly modifying the many-body properties of the system. 
These effects are seen both in the ground state and elementary excitations, which are governed by a delicate balance between the intra- and inter-layer interactions.
In a recent work~\cite{filinovPRA16}, the elementary excitation spectrum at finite temperature of an arrangement of dipoles in a bilayer geometry has been analyzed.

In the present work, we perform a microscopic calculation of the dynamic response of the bilayer system at zero temperature.
Our results rely on the use the diffusion Monte Carlo (DMC) method in combination with the correlated basis function (CBF) theory.
The CBF-DMC combination has proved useful for dipolar quantum gases~\cite{Mazzanti09,Macia_2012}, but also for molecule dynamics in
$^4$He droplets~\cite{zillichPRB04,zillichPRL04,zillichJCP08,zillichJCP10}.
We consider a system of dipoles with moments oriented perpendicularly to the two planes where they are allowed to move.
The bilayer setup offers the somehow unique opportunity of realizing a Bose gas with a gapped spectrum, so we pay special attention to the spin-channel excitations.

The rest of the paper is organized as follows.
In Section~\ref{sec:method}, we introduce the model and the methods used in the study, i.e., the DMC method and the CBF theory adapted to the bilayer geometry.
Results for the density and spin dynamic responses are reported in Sec.~\ref{sec:results}, with special attention to the development of a gap in the spin channel of the pair superfluid regime.
Finally, Sec.~\ref{sec:conclusions} comprises a brief summary of the obtained results and the main conclusions of our work.

\section{Method\label{sec:method}}

The bilayer system under study is described by a two-component Hamiltonian with bosonic dipoles in two two-dimensional parallel layers $A$ and $B$, separated by a distance $h$,
\begin{eqnarray}
H&=&-\frac{\hbar^2}{2m}\sum_{\alpha=A,B}\sum_{i=1}^{N_\alpha}\nabla_{i,\alpha}^2
\\
&+&
\sum_{i<j}^{N_A}\frac{d^2}{r_{iA,jA}^3} + \sum_{i<j}^{N_B} \frac{d^2}{r_{iB,jB}^3}
+ \sum_{i,j}^{N_A,N_B}
\frac{d^2(r_{iA,jB}^2-2h^2)}{(r_{iA,jB}^2+h^2)^{5/2}}
\;. \nonumber
\label{hamiltonian}
\end{eqnarray}
The first line describes the kinetic energy of particles of mass $m$ (equal in both layers) and the second line corresponds to the intra-layer and inter-layer dipolar interactions. 
Each layer contains $N_A=N_B=N/2$ dipoles, with the dipole moment $d$ oriented perpendicularly to the layers.
Here, $r_{i\alpha,j\beta}=|{\bf r}_{i\alpha}-{\bf r}_{j\beta}|$ denotes the distance between particle $i$ in layer $\alpha$ and particle $j$ in layer $\beta$.
For $\alpha=\beta$ (same layers) the distance is the in-plane distance, while for $\alpha\ne\beta$ it is the distance between the projections onto any of the layers of the positions of the $i$-th and $j$-th particles.

\subsection{Diffusion Monte Carlo method}

The ground-state properties of the system, described by the Hamiltonian~(\ref{hamiltonian}), can be efficiently calculated by means of the diffusion Monte Carlo (DMC) method.
DMC solves stochastically the $N$-body Schr\"odinger equation in imaginary time and provides not only the energy and static properties, but also correlation functions in imaginary time.
For a Bose system like the present one, DMC is an exact method, constrained by statistical noise which, on the other hand, can be accurately estimated.
We use the same number of particles, time steps, guiding wave function, and other technical parameters as in Ref.~\cite{Macia_2014}.

Our main goal is the study of the excitations of the bilayer system.
Therefore, we use the DMC method to calculate the intermediate scattering functions in imaginary time since real time dynamics is not accessible,
\begin{eqnarray}
  S_{\alpha\beta}({\bf k},\tau) = \frac{1}{N} \langle \rho_\alpha ({\bf k}, \tau)
  \rho_\beta (-{\bf k}, 0) \rangle \ ,
\label{eq:S(k,tau)}
\end{eqnarray}
where $\rho_\alpha({\bf k}, \tau) = \sum_{j} \exp(i {\bf k} \cdot {\bf r}_{j\alpha}(\tau))$ is the density operator for particles in layer $\alpha$, in the momentum representation.
The values of the intermediate scattering functions at $\tau=0$ are the corresponding static structure factors, $S_{\alpha \beta}({\bf k})$.

The dynamic structure function matrix $S_{\alpha\beta}({\bf k},\omega)$ is related to the imaginary-time function~(\ref{eq:S(k,tau)}) by the Laplace transform
\begin{eqnarray}
  S_{\alpha\beta}({\bf k},\tau) = \int\limits_{0}^\infty e^{-\tau\omega}
                                S_{\alpha\beta}({\bf k},\omega) \, {\rm d}\omega \;.
\label{eq:S(k,omega)}
\end{eqnarray}
In our case of identical bilayers, $S_{AA}({\bf k},\omega)=S_{BB}({\bf k},\omega)$ and
$S_{AB}({\bf k},\omega)=S_{BA}({\bf k},\omega)$. $S_{\alpha\beta}({\bf k},\omega)$ is diagonalized to obtain the density response $S_s({\bf k},\omega)=S_{AA}({\bf k},\omega)+S_{AB}({\bf k},\omega)$, which probes the symmetric mode, where particles in both layers move in phase; and the ``spin response'' $S_a({\bf k},\omega)=S_{AA}({\bf k},\omega)-S_{AB}({\bf k},\omega)$
which probes the antisymmetric mode, where particles in both layers move out of phase.

As it is well known, calculation of the inverse Laplace transform required to obtain the dynamic response from the intermediate scattering function $S_{\alpha\beta}({\bf k},\tau)$ is a mathematically ill-conditioned problem.
At the practical level, this means that the always-present finite accuracy of the input data makes it impossible to find a unique reconstruction of the dynamic structure factor.
Among the techniques devised specifically to get a better signal for the dynamic response we used a stochastic optimization approach.
In particular, we work with a simulated annealing method that recently has proved to be a reasonable approach to study dynamics in a quantum many-particle system~\cite{ferre}.

It was shown in Ref.~\cite{Macia_2014} that a gap opens up in the pair-superfluid phase of the bilayer, and thus a finite energy is needed to create an excitation in the spin mode, i.e.\/ the mode where the partial densities fluctuate out of phase in the two layers, see Sec.~\ref{sec:cbf}.
The value of the gap~$\Delta$ can be extracted from the ground-state energy as the difference of chemical potentials between the $N+1$ and the $N$ particle system,
\begin{equation}
2\Delta = E_{N+1} - 2 E_{N} + E_{N-1}
\;.
\label{Eq:gap:E}
\end{equation}
where $E_{N+1}$ is the energy of a bilayer with an additional particle in one of the layers and $E_{N-1}$ is the energy with one particle less in one of the layers.
The gap is related to the difference in the energy between odd and even number of particles.
A value $\Delta \ne 0$ means that the system has a pairing energy, as in superfluid Fermi gases.
In the limit of small distances between layers, this energy gap becomes large and equal to the binding energy $E_b$ of a dimer composed by an upper and a lower dipole, $2\Delta = E_b$.

\subsection{Correlated basis function theory for a bilayer of bosons \label{sec:cbf}}

The CBF method for layers of finite thickness was introduced in Ref.~\cite{clementsPRB96},
and applied to dipolar Bose condensates~\cite{hufnaglPRL11,hufnaglPRA13}.
In the present work, we are interested in a two-component Bose gas.
The multi-component CBF theory was derived in Ref.~\cite{raderInprogress}.
CBF relies on a time-dependent version of the Jastrow-Feenberg wave function
\begin{eqnarray}
  \psi(t)\sim e^{-i E_0 t} e^{\frac{1}{2} \excitationOp(t)} \psi_0
  \ \text{,}
\end{eqnarray}
assuming that the ground-state energy $E_0$ and the averages over the ground-state wave
function $\psi_0$ are known, in the present case from DMC simulations.
The excitation operator is
\begin{eqnarray}
  \excitationOp(t) = \sum\limits_{\alpha, j} \deltau_\alpha(\rv_{j\alpha})
    + \frac{1}{2} \sumprime\limits_{\alpha, \beta, j, k}
      \deltau_{\alpha \beta}(\rv_{j\alpha}, \rv_{k\beta}, t)
\label{eq:deltaU}
\end{eqnarray}
with the one- and two-body correlation fluctuations $\deltau_\alpha$ and $\deltau_{\alpha \beta}$.
The prime in the second sum means that terms $j = k$ are excluded if $\alpha = \beta$.
Equations for $\deltau_\alpha$ and $\deltau_{\alpha \beta}$ are obtained using the minimum action principle, which is equivalent to solving the time-dependent Schr\"odinger equation
\begin{eqnarray}
  \delta\int\! \diff{t}\ \langle \psi(t) \vert \hamiltonian(t) - i \hbar \partial_t \vert \psi(t) \rangle = 0
\ ,
\label{eq:deltaS}
\end{eqnarray}
where $\hamiltonian(t) = \hamiltonian_0 + \hamiltonian_1(t)$ is the many-body Hamiltonian perturbed by an arbitrary one-body potential
\begin{eqnarray}
  \hamiltonian_1(t) = \sum_{\alpha} \sum\limits_{j}^{N_\alpha}
  \potential_{\alpha}(\rv_{j\alpha}, t)
\ \text{.}
\end{eqnarray}
We are interested in the response of the system to a weak perturbation such that the  Euler-Lagrange equations resulting from Eq.~(\ref{eq:deltaS}) can be linearized.
Using further assumptions (uniform limit approximation and the convolution approximation~\cite{raderInprogress}), we arrive at the linear relation between the perturbation $\potential_{\alpha}$ and the density fluctuation $\Delta\rho_\alpha$ which defines the density response operator $\chi$.
For a perturbation with wavenumber $\kv$ and frequency $\omega$ we obtain
\begin{eqnarray}
  \Delta\rho_\alpha(k,\omega)
= \sum_\beta \chi_{\alpha \beta}(k,\omega)\, V_\beta(k,\omega)
\ \text{.}
\end{eqnarray}
The density response operator $\chi$ in the CBF approximation is given by
\begin{eqnarray}
  \chi_{\alpha \beta}(k,\omega)
&=& \sqrt{\rho_\alpha\rho_\beta}\sum_{m,n} \phi_{n,\alpha}(k)\phi_{m,\beta}^*(k) \big[ G_{mn}(k,\omega)\nonumber\\
&& \qquad\qquad +  G_{mn}^*(k,-\omega)\big]
  \ \text{,}
\end{eqnarray}
where $G_{mn}$ is defined by the inversion of
\begin{eqnarray}
  G_{mn}^{-1}(\kv,\omega) =
  \big(\hbar \omega - \varepsilon_m(k) + i\eta\big)\delta_{mn} - \Sigma_{mn}(k,\omega)
\ \text{.}
\end{eqnarray}
$\varepsilon_m(k)$ is the Bijl-Feynman approximation to the excitation energies of momentum $\hbar\kv$, where $m=1,2$ numbers the two modes of the coupled two layers.
The energy $\varepsilon_m(k)$ is obtained by solving
\begin{eqnarray}
  {\hbar^2 k^2\over 2m} \psi_{m,\alpha}(k) = \varepsilon_m(k) \sum_\beta S_{\alpha\beta}(k)\psi_{m,\beta}(k)
  \ \text{,}
\end{eqnarray}
where $S_{\alpha\beta}(k)$ is the matrix of static structure factors mentioned above.
Since the two layers are identical, the two solutions to this generalized $2\times 2$
eigenvalue problem are the symmetric mode $s$ and the anti-symmetric mode $a$ mentioned
above, with eigenvectors $(\psi_{s,\alpha}(k))_{\alpha}\sim (1,1)$ and
$(\psi_{a,\alpha}(k))_\alpha\sim (1,-1)$, respectively~\cite{hebenstreitPRA16}.
The corresponding Bijl-Feynman energies are
\begin{eqnarray}
\varepsilon_s(k)&=&{\hbar^2k^2\over 2m}(S_{AA}(k)+S_{AB}(k))^{-1}\label{eq:epsilons}\\
\varepsilon_a(k)&=&{\hbar^2k^2\over 2m}(S_{AA}(k)-S_{AB}(k))^{-1}\label{eq:epsilona}
\ \text{.}
\end{eqnarray}
In the symmetric mode (density mode), particles in the two layers oscillate in phase leading to an oscillation of the total density, $\Delta\rho_A+\Delta\rho_B\ne 0$.
In the antisymmetric mode (spin mode), particles in the two layers oscillate out of phase, such that $\Delta\rho_A+\Delta\rho_B = 0$.
The Bijl-Feynman approximation is obtained from~(\ref{eq:deltaS}) by discarding two-body correlation fluctuations $\deltau_{\alpha \beta}$ in the excitation operator~(\ref{eq:deltaU}).
Note that the Bijl-Feynman approximation predicts for both modes a linear dispersion in the long wave length limit $k\to 0$, as long as $S_{AA}(k)$ and $S_{AB}(k)$ have a different slope for $k\to 0$.

The self energy $\Sigma_{mn}$ results from the inclusion of two-body correlation fluctuations $\deltau_{\alpha \beta}$.
The CBF expression for $\Sigma_{mn}$ can be found in Ref.~\cite{raderInprogress}.
It can be interpreted as corrections due to coupling of Feynman excitation modes, leading to an overall reduction of excitation energies, but also to damping.
For a symmetric arrangement, it can be shown that $\Sigma_{mn}=0$ if $m-n$ is odd.
Since $\Sigma_{mn}$ is a $2\times 2$ matrix in the present case, it is diagonal.
Thus we get the same modes $m=s,a$, symmetric and antisymmetric, as in the Feynman approximation, but modified by the self energy, $\Sigma_{ss}$ or $\Sigma_{aa}$.
We can thus define two density response function, $\chi_s(k,\omega)$ for the symmetric mode and $\chi_a(k,\omega)$ for the anti-symmetric mode
\begin{eqnarray}
&\chi_s(k,\omega) =G_{ss}(k,\omega)+  G_{ss}^*(k,-\omega)\\
&\chi_a(k,\omega) =G_{aa}(k,\omega)+  G_{aa}^*(k,-\omega)
\end{eqnarray}
The dynamic structure function $S_{s(a)}(k,\omega)$ for the symmetric (antisymmetric) modes is the imaginary part of $\chi_{s(a)}(k,\omega)$, according to the fluctuation dissipation theorem.
Since the imaginary part of $G_{nn}$ vanishes for negative frequencies, one obtains
\begin{eqnarray}
&S_s(k,\omega) ={1\over\pi}\Im\left[ \hbar \omega - \varepsilon_s(k) + i\eta - \Sigma_{ss}(k,\omega) \right]^{-1} \\
&S_a(k,\omega) ={1\over\pi}\Im\left[ \hbar \omega - \varepsilon_a(k) + i\eta - \Sigma_{aa}(k,\omega) \right]^{-1}
\ \text{.}
\end{eqnarray}

\section{Results\label{sec:results}}

Using the methods discussed in Sec.~\ref{sec:method} we calculate the dynamic structure function of a bilayer geometry, focusing on the nature of the excitations when the system changes from the atomic phase (single superfluid) to the dimer one (pair superfluid).
All the results have been calculated at a density $nr_0^2 = 1$, which corresponds to an intermediate value where dipolar effects are already strong, but where the system still remains in a gas phase~\cite{Macia_2014}.
We use $r_0=md^2/\hbar^2$ and $E_0=\hbar^2/(mr_0^2)$ as natural units for length and energy, respectively.
We consider two characteristic values of the interlayer spacings, $h = 0.4 r_0$ and $h = 0.2r_0$.
For the larger spacing the system remains in an atomic phase and for the smaller $h$ the interlayer attraction leads to the formation of dimers, i.e.\/ bound states of two dipoles from different layers.

\begin{figure}[t!]
\begin{center}
\includegraphics*[width=0.48\textwidth]{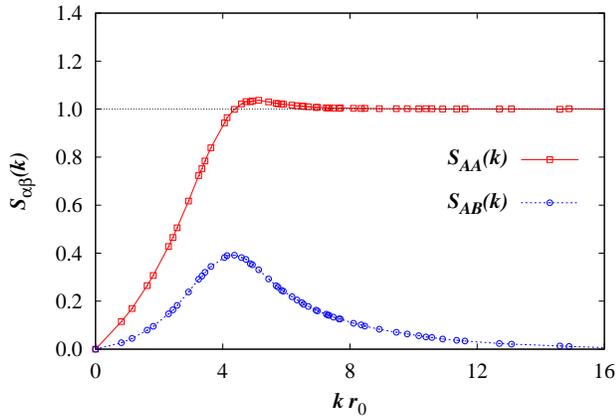}
\end{center}
\caption{(color online)
The diagonal and off-diagonal parts, $S_{AA}(k)$ and $S_{AB}(k)$ of the static structure function matrix of the ground state of two dipolar layers separated by $h=0.4r_0$.
Error bars are smaller than the symbol size.
}
\label{Fig:s4}
\end{figure}

\begin{figure}[t!]
\begin{center}
\includegraphics*[width=0.48\textwidth]{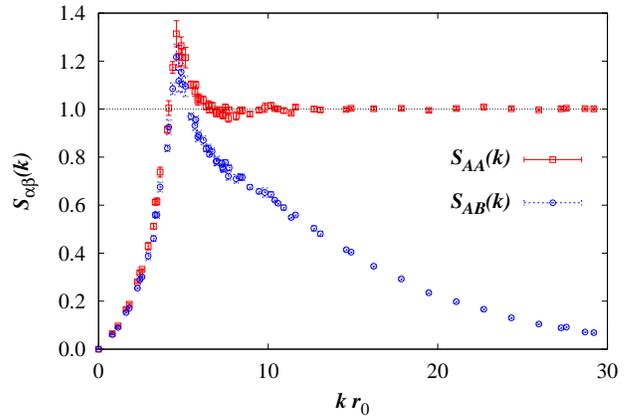}
\end{center}
\caption{(color online)
Same as Fig.~\ref{Fig:s4} for $h=0.2r_0$.
}
\label{Fig:s2}
\end{figure}

Before discussing the evolution of the dynamics as a function of $h$, we report results for the static structure factors, $S_{AA}(k)$ and $S_{AB}(k)$, obtained from DMC simulations using the pure estimator~\cite{Casulleras95}.
Figures~\ref{Fig:s4} and~\ref{Fig:s2} show $S_{AA}(k)$ and $S_{AB}(k)$ for the two layer separations considered, $h=0.4r_0$ and $h=0.2r_0$, respectively.
In the absence of any interlayer coupling, the off-diagonal element of the static structure function matrix vanishes, $S_{AB}(k)=0$.
In the opposite limit, $h\to 0$, pairs of dipoles are tightly locked together producing effectively a single-component Bose gas of dimers.
In this limit, $S_{AB}(k) \to S_{AA}(k)$.
For $h=0.4r_0$ (Fig.~\ref{Fig:s4}) the coupling between dipoles in the two layers is already quite strong, but not enough for dimerization (see Ref.~\cite{Macia_2014}); indeed $S_{AB}(k)$ is well below $S_{AA}(k)$ for all $k$.  For $h=0.2r_0$ (Fig.~\ref{Fig:s2}), i.e.\/ in the dimerized case, we see that $S_{AB}(k)\approx S_{AA}(k)$ up to the peak located at $kr_0\approx 4$.
Only for larger $k$ values, $S_{AB}(k)$ starts to fall below $S_{AA}(k)$, eventually decaying to zero.
In other words, the pair superfluid phase, where the dimers can be considered as individual particles, is observed to emerge in the mixed static structure factor $S_{AB}(k)$ at low and intermediate $k$ values.

\begin{figure}[t!]
\begin{center}
\includegraphics*[width=0.45\textwidth,angle=0]{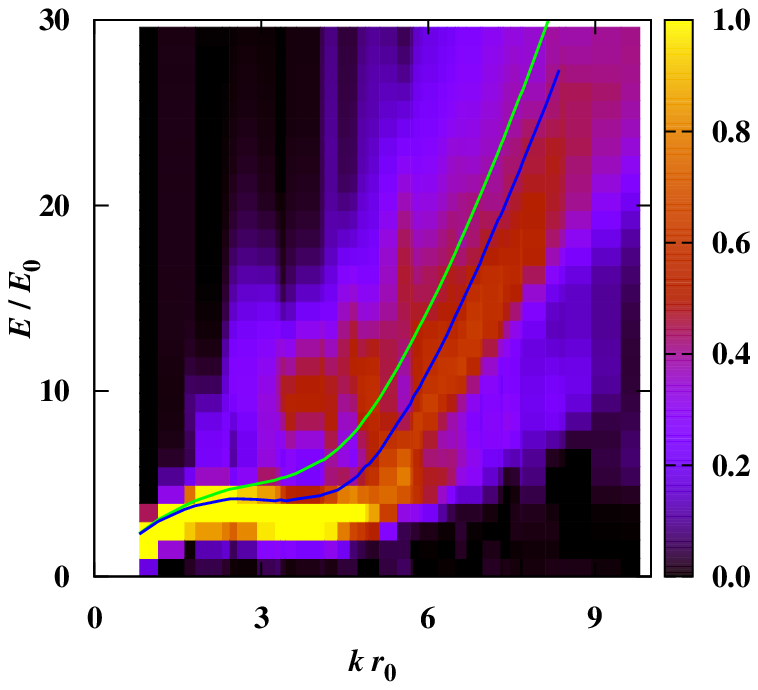}
\includegraphics*[width=0.45\textwidth,angle=0]{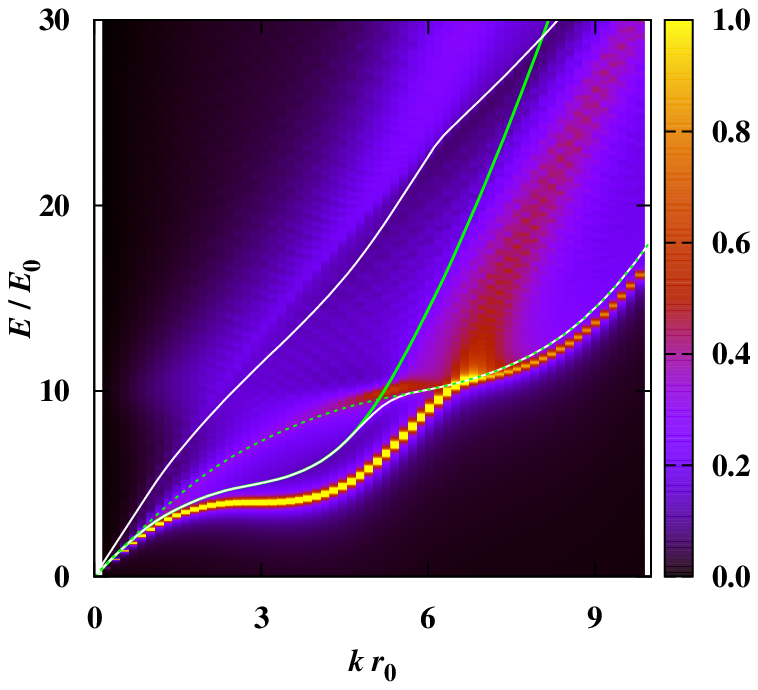}
\end{center}
\caption{(color online)
Color map of the symmetric (density) response $S_s(k,\omega)$ in the atomic phase (interlayer spacing $h=0.4r_0$) as obtained from
(a) the inverse Laplace transform (top panel) and (b) the CBF-DMC method (bottom panel).
In order to highlight broad features with low amplitude, we mapped the square root of $S_s(k,\omega)$ to the given color scale.
The green lines are the Feynman upper bound~(\ref{eq:epsilons}).
The lower blue line in (a) is a single-exponent fit~(\ref{Eq:single exponent}).
In (b), the dashed line is the energy of two Bijl-Feynman excitations at half the wavenumber $k$, $2\varepsilon_{s/a}(k/2)$,
and the white lines are the dissipation borders $b_s^{(s)}(k)$ and $b_s^{(a)}(k)$, Eq.~(\ref{eq:borders}).
}
\label{Fig:density:0.4}
\end{figure}

Figure~\ref{Fig:density:0.4} shows the dynamic structure function $S_s(k,\omega)$ for the density response (the symmetric mode) in the atomic phase at $h=0.4r_0$, as obtained from the inverse Laplace transform (top) and CBF theory using the DMC results of the static structure factors as input (bottom).
For small momenta, the spectrum is exhausted by a single branch of excitations, which has a linear phononic dispersion relation, $E(k) = \hbar k c_s$, with $c_s$ the speed of sound.
Since the Bijl-Feynman approximation works well for long wave lengths, we can obtain $c_s$ from the $k\to 0$ limit of Eq.~(\ref{eq:epsilons}), $c_s={\hbar\over 2m}(S'_{AA}+S'_{AB})^{-1}$, with the slopes $S'_{\alpha\beta}=dS_{\alpha\beta}(k)/dk|_{k=0}$.
Alternatively, $c_s$ could be determined from the generalization of the relation between the compressibility and the speed of sound.
In a single-component system, $mc_s^2 = \rho d^2e / d\rho^2$, where $e$ is the energy density (energy per volume).
In a symmetric binary system (i.e. equal partial densities $\rho_\alpha=\rho/2$) this relation is generalized to $mc_s^2=\rho(e_{AA}-e_{AB})/2$, with the second derivatives with respect to partial densities $e_{\alpha\beta}=\partial^2 e / \partial\rho_\alpha\partial\rho_\beta$, see e.g. Ref.~\cite{campbellJLTP71}.

At larger momenta, a roton minimum starts getting formed (for higher densities the minimum becomes more evident~\cite{Mazzanti09,filinovPRA16}).
Here, the Bijl-Feynman approximation $\varepsilon_m(k)$ lies above the lower branch, as the high-energy excitations provide important contributions.
In the CBF approximation, the dynamic structure function for the density response has a rich structure in the atomic phase.
Several branches of excitations are resolved while the inverse Laplace transform provides a lower-quality picture, with only the most intense branch resolved.
Both methods make it evident that the Bijl-Feynman approximation is precise only for low momenta, while for higher $k$ it predicts excitation energies that are too high, being an upper bound.
In particular, for the larger interlayer separation $h=0.4r_0$ the Bijl-Feynman dispersion has a positive slope everywhere.
Instead, the correct result is that a roton starts to form in the density response, as it can be seen from the inverse Laplace method and even better from the CBF-DMC approach.
Up to about $kr_0=6$, most of the spectral weight of $S_s(k,\omega)$ is carried by a phonon-roton spectrum.

\begin{figure}[t!]
\begin{center}
\includegraphics*[width=0.45\textwidth,angle=0]{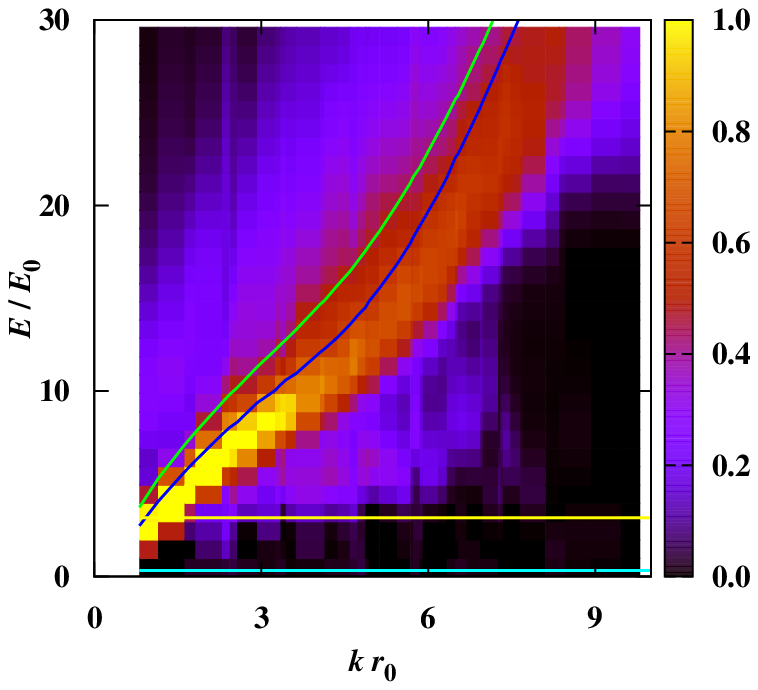}
\includegraphics*[width=0.45\textwidth,angle=0]{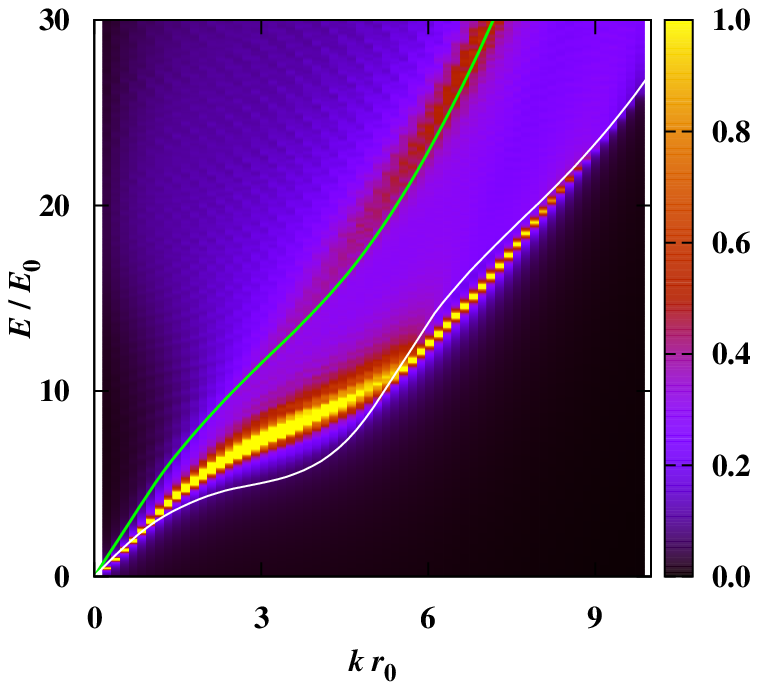}
\end{center}
\caption{(color online)
Color map of the antisymmetric (spin) response $S_a(k,\omega)$ in the atomic phase (interlayer spacing $h=0.4r_0$).
The notation is the same as in Fig.~\ref{Fig:spin:0.4}, the white line now showing the
dissipation borders $b_a(k)$, Eq.~(\ref{eq:bordera}).
The upper yellow line in (a) denotes the dimer binding energy $E_b$
Twice the gap energy $2\Delta$, Eq.~(\ref{Eq:gap:E}),
shown by the lower cyan line, is negligible.
}
\label{Fig:spin:0.4}
\end{figure}

The white lines in the map of the CBF-DMC result for the symmetric dynamic structure function indicate the dissipation borders above which the decay of a symmetric excitation with momentum $k$ into two symmetric (lower line) or two anti-symmetric modes (upper line) is kinematically allowed,
\begin{eqnarray}
b_s^{(s/a)}(k)={\rm min}_q[\varepsilon_{s/a}(q)+\varepsilon_{s/a}(|\kv-\qv|)]\;.
\label{eq:borders}
\end{eqnarray}
Excitations below the dissipation border for decay into two symmetric modes are therefore undamped, as is the case for the phonons and rotons; their finite width in Fig.~\ref{Fig:density:0.4} comes from an artificial Lorentzian broadening ($0.07E_0$) used for plotting the response.  The effect of the dissipation border for decay into two antisymmetric modes is less dramatic, because it only marks an {\em additional} decay process.
We note that due to the approximations made in the derivation of the CBF method, the decay happens into Bijl-Feynman modes, see Ref.~\cite{raderInprogress}.

Around $kr_0 \approx 6.5$, the phonon-roton dispersion crosses the dissipation border $b_s^{(s)}(k)$ and splits into a strongly damped mode and weaker undamped mode slightly below the border.
We also show the energy of two Bijl-Feynman excitations, each with half the wavenumber $k$, $2\varepsilon_{s/a}(k/2)$, as a thin dashed line.  For a $k$-range around the roton, $S_s(k,\omega)$ indeed has some spectral strength for $2\varepsilon_{s/a}(k/2)$.
This indicates that there is a non-negligible response to a perturbation of wavenumber $k$ where two modes, each with momentum $k/2$, are simultaneously excited.
The reason for that is the high density of states where the dispersion has a small slope.

Above the border for decay into two antisymmetric spin modes, the CBF-DMC density response exhibits additional structure.
In particular we observe broad, strongly damped dispersion which, for even higher energies, eventually attains a free particle spectrum.
We want to stress that the details of the dynamics response function are expected to depend on the approximations made within CBF, and may change if improved theories are used~\cite{campbellPRB09,campbellJLTP10,campbellPRB15}.

If the lower branch has a high intensity, a simple one-exponent fit to the imaginary-time dynamic structure function
\begin{eqnarray}
S(k,\tau) = Z \exp(-\omega(k) \tau )
\label{Eq:single exponent}
\end{eqnarray}
is able to capture its position. We find that the single-exponent method provides a reasonable agreement for the position of the lower branch up to momenta $k r_0 \approx 7$.
As discussed above, the structure of excitations changes for larger momenta and a single-mode description is no longer applicable.

Figure~\ref{Fig:spin:0.4} reports the spin response $S_a(k,\omega)$ (the antisymmetric mode) in the atomic phase.
The lower branch is clearly visible and it is linear for small momenta.
Similarly to the density mode, in the Bijl-Feynman approximation the speed of the spin wave can be obtained from Eq.~(\ref{eq:epsilona}), $c_a={\hbar\over 2m}(S'_{AA}-S'_{AB})^{-1}$.
Importantly, in the limit of zero momentum the excitation energy vanishes in the CBF-DMC spectrum, so there is no gap in the spin sector. 
The spin response obtained from the inverse Laplace transform is compatible with this result, although $k$ is bounded from below due to the finite size of the simulation box.
Also the binding energy of dimers shown in Fig.~\ref{Fig:spin:0.4} does not play any role in $S_a(k,\omega)$.
Again, the CBF-DMC approach provides more detailed structure compared to the inverse Laplace transform.
For the antisymmetric mode, there is only one dissipation border, namely for decay into a symmetric and an antisymmetric mode, indicated by the white line in Fig.~\ref{Fig:spin:0.4},
\begin{eqnarray}
b_a(k)={\rm min}_q[\varepsilon_{s}(q)+\varepsilon_{a}(|\kv-\qv|)]\;.
\label{eq:bordera}
\end{eqnarray}
In this way, the spin response $S_a(k,\omega)$ has a simpler structure than the density response $S_s(k,\omega)$.
For intermediate momenta, most of the weight is carried by an excitation well below the Bijl-Feynman approximation.
The spin mode is above the dissipation border for wavenumbers up to about $kr_0\approx 5.5$ and therefore it is damped.
But beyond $kr_0\gtrsim 5.5$, the dispersion exits the dissipative regime by going below the dissipation border and becomes undamped until it crosses the border again around $kr_0\approx 9$.
Although at such high $k$ value, significant spectral weight has been shifted to a high-energy mode, which becomes the free particle mode for very large $k$, a window of wavenumbers for long-lived antisymmetric excitations of a dipolar bilayer is of experimental relevance.
The wavy pattern visible in Fig.~\ref{Fig:spin:0.4} are numerical artifacts due to finite discretization.

\begin{figure}[t!]
\begin{center}
\includegraphics*[width=0.45\textwidth,angle=0]{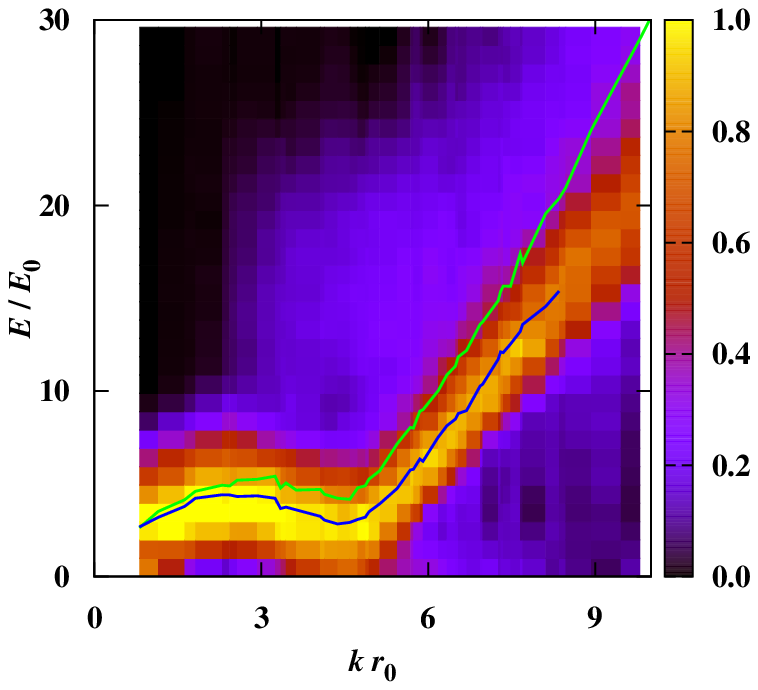}
\includegraphics*[width=0.45\textwidth,angle=0]{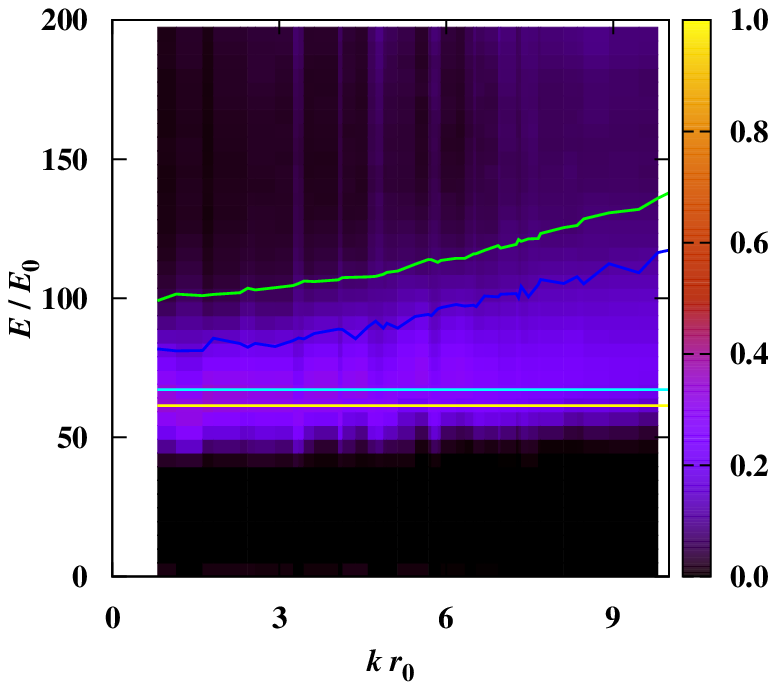}
\end{center}
\caption{(color online)
Color maps of the dynamics structure functions for a layer spacing of $h=0.2r_0$, i.e.\/ in the dimer phase, obtained from the inverse Laplace transform.
The upper panel shows the symmetric/density response and the lower one the antisymmetric/spin response.
The notation is the same as in Figs.~\ref{Fig:density:0.4} and~\ref{Fig:spin:0.4}.
}
\label{Fig:0.2}
\end{figure}

For the small interlayer distance, $h=0.2r_0$, dipoles from different layers are locked into a dimer, i.e.\/ a bound state.
Dimerization is not accounted for in the multi-component CBF generalization outlined in section~\ref{sec:cbf} and therefore it cannot be applied for such small $h$.
Nevertheless, the inverse Laplace method can be used to predict the characteristic features of the response functions.
Figure~\ref{Fig:0.2} shows the dynamic structure factor for the density and spin channels.
In the density channel (top panel), a strong roton is observed, manifesting much stronger correlations between the particles.
The reason is that a dimer features twice the atom mass and dipole moment as compared to single dipoles.
The net effect is an increase in the dimensionless (total) density from $nr_0^2=1$ for atoms to the effective
dimer density $nr_0^2 =32$~\cite{Macia_2014}, at which the roton is well formed~\cite{Astrakharchik07a}.
The main differences are observed in the spin response, where the structure changes dramatically.
One can see how a gap has opened in the excitation spectrum.
A finite energy is needed to create (antisymmetric) excitations in the spin channel, as a finite pairing energy has to be expended.
We verify in Fig.~\ref{Fig:0.2} that the energy which is needed is exactly $2\Delta$, with $\Delta$ obtained by the staggering method defined by Eq.~(\ref{Eq:gap:E}).
This energy is similar to the binding energy of a dimer $2\Delta \approx E_b$, and both quantities coincide in the limit of small interlayer distances, $h\to 0$.
Recent simulations performed using the path integral Monte Carlo method have shown that the gap closes as the temperature is increased~\cite{filinovPRA16}.
Finally, we note that there is a large separation of scales in the molecular case, with the spin excitations of the order of the binding energy (highly energetic) and the density excitations of the order of phonon energies (low energies).

\section{Conclusions\label{sec:conclusions}}

A bilayer of dipolar bosons is a unique setup permitting to study the continuous transition between an atomic Bose superfluid and a pair superfluid in a controlled way.
By adjusting the interlayer distance between the layers one can determine with high precision and tunability the evolution between both regimes.
In previous work~\cite{Macia_2014}, this transition was characterized relying on ground state properties such as the energy, the condensate fraction of atoms/dimers, and the superfluidity through the calculation of winding numbers.
In the present work, we address this transition by looking at the excitations of the system on both sides of it.

Our main goal has been the calculation of the dynamic response function, which contain the maximum attainable information on the excited states of the system.
As we deal effectively with a two-component system, consisting of dipoles on the top and bottom layers, the most relevant physical information is contained in the symmetric (density) and antisymmetric (spin) components of the dynamic response.
However, the estimation of these quantities is much more difficult than the ground-state properties since quantum Monte Carlo methods are designed to arrive to the ground state by propagating the system in imaginary time.
In principle, from the calculation of the intermediate scattering functions in imaginary time one can obtain the dynamic structure functions through an inverse Laplace transform.
In practice, this inverse problem is ill-conditioned for the noisy data obtained from the simulations, and it is not possible to arrive to an unambiguous optimal solution.
In order to tackle this severe drawback, we adopted two approaches.
In the first one, we use CBF theory using as inputs the ground-state static structure factors from DMC.
The CBF-DMC combination was used in the past, providing an excellent description of the excited states of an $N$-body problem~\cite{Macia_2012,Mazzanti09}.
Unfortunately, the present multi-component CBF theory works only for the atomic phase, not the dimerized phase.
A second approach is the numerical reconstruction of the dynamic response from the imaginary-time intermediate scattering functions using a multidimensional optimization method, namely simulated annealing.
This method has been recently used in the calculation of the dynamic response in liquid $^4$He with reasonable success~\cite{ferre}.
The output of this second approach is significantly broader than the CBF-DMC one but has the advantage of being applicable also to the pair-superfluid regime.
It is worth mentioning that a similar optimization method was used recently by Filinov in the study of the bilayer at finite temperature using path integral Monte Carlo data~\cite{filinovPRA16}.

Our results show unambiguously the change in the nature of the excitations when the system evolves from an atomic regime to a dimerized one by decreasing the interlayer distance $h$.
In the atomic phase, with single-atom superfluidity, we observe an low-energy spectrum of phonon-type, both in the density and spin channels. In particular, the spin response goes linearly to zero when $k \to 0$.
The energy spectrum for the density response has a roton, albeit a shallow one.
The description changes dramatically when stable dimers (pair superfluid) form the ground-state configuration of the system. 
The change in the density mode is essentially quantitative; it is still of phonon-roton type as expected, but with a much deeper roton minimum.
The observation of rotons in dilute gases has been widely discussed and several proposal were made~\cite{Santos_2003,Esslinger}.
Probably, one of the best setups for observing a significant roton in dilute systems would be the use of bilayer or multilayer stacks of dipolar bosons to produce effective dipolar moments much larger than in a single two-dimensional trap.

The most dramatic change in the response upon dimerization is observed in the spin mode.
In the pair-superfluid regime our calculations show unambiguously the presence of a gap of high energy, which for small interlayer separation coincides with half the binding energy of the dimer.
Whereas the existence of a gap in a superfluid Fermi system due to the pairing mechanism is well known and understood, the observation of a gap mode in a Bose gas is noticeable.
Hopefully, in the near future it will be possible to design bilayer setups with a tunable interlayer distance or with bosons with tunable dipole moments which can reach the pair-superfluid regime and, through Bragg scattering~\cite{Landig}, observe the predicted gap.

\begin{acknowledgments}

This work was supported by the Austrian Science Fund FWF under Grant P23535 and the MICINN (Spain) Grant FIS2014-56257-C2-1-P.
The Barcelona Supercomputing Center (The Spanish National Supercomputing Center - Centro Nacional de Supercomputaci\'on) is acknowledged for the provided computational facilities.

\end{acknowledgments}

\end{document}